\documentclass{article}
\usepackage{amsmath,amsfonts,amssymb,upmath}
\def\bm#1{\mathchoice
 {\mbox{\boldmath$\displaystyle#1$}}%
 {\mbox{\boldmath$#1$}}%
 {\mbox{\boldmath$\scriptstyle#1$}}%
 {\mbox{\boldmath$\scriptscriptstyle#1$}}}

\begin{document}

 \title{Coarse-grained~description of~a~passive~scalar}
 \author{Antonio Celani\thanks{CNRS, INLN, 1361 Route des Lucioles, 06560 Valbonne, France}, Marco Martins Afonso$^{\ddagger}$\thanks{Department of Physics of Complex Systems, The Weizmann Institute of Science, Rehovot 76100, Israel} \& Andrea Mazzino\thanks{Department of Physics - University of Genova \& INFN - Genova Section, via Dodecaneso 33, 16146 Genova, Italy}}
 \date{\today}
 \maketitle

\begin{abstract}
 The issue of the parameterization of small-scale dynamics is addressed in the context of passive-scalar turbulence. The basic idea of our strategy is to identify dynamical equations for the coarse-grained scalar dynamics starting from closed equations for two-point statistical indicators. With the aim of performing a fully-analytical study, the Kraichnan advection model is considered. The white-in-time character of the latter model indeed leads to closed equations for the equal-time scalar correlation functions. The classical closure problem however still arises if a standard filtering procedure is applied to those equations in the spirit of the large-eddy-simulation strategy. We show both how to perform exact closures and how to identify the corresponding coarse-grained scalar evolution.
\end{abstract}

\section{Introduction}

 A common feature characterizing turbulent systems is the presence of space and time fluctuations \cite{F95}. Depending on the strength of some control parameters (e.g. the Reynolds number in hydrodynamic turbulence or the P\'eclet number in scalar turbulence) the number of active degrees of freedom can reach values arbitrarily large and eventually diverge in the limit of the so-called fully developed turbulence regime \cite{F95}.\\
 If, on the one hand, this huge number of active scales characterizing the latter regime is the ideal framework to investigate `classical' problem in the realm of basic turbulent research (e.g. in relation to global scale invariance and its violation in the form of intermittency and anomalous scaling), on the other hand the proliferation of degrees of freedom leaves severe limitations to the deterministic description of turbulent fields. To have a quantitative idea, in the turbulent atmospheric boundary layer, the degrees of freedom to describe all active scales can reach values of the order of, or larger than, $10^{27}$ ($10^9$ in each spatial direction \cite{L97}). This number is obviously far from being handled even by state-of-the-art super computers.\\
 As a matter of fact, in many situations of practical interest (e.g. the description of the evolution of a pollutant emitted by sources in the atmospheric boundary layer) one is not interested in describing the details (i.e. small-scale dynamics) of turbulent fields but, rather, in focusing on their large-scale behaviour. This is actually only an apparent simplification. Indeed, the nonlinear character of many turbulent systems (e.g. hydrodynamic turbulence) does not permit to restrict the attention on the sole (large) scales of interests: small scales enter large-scale dynamics in a way that the latter is not closed. This is the well-known closure problem \cite{MK00}. It is worth stressing that closure problems also appear in the absence of nonlinearity. The best example is provided by passive-scalar turbulence, where the closure problem arises from the advection term (i.e. a term involving a multiplicative structure between velocity and scalar).\\
 The present situation, in relation to the long-standing closure problem, is that a unifying theory predicting how to close the basic equations in the large scales is not available. Readers can refer to \cite{MK00} for a modern review on the closure problem oriented to the so-called large-eddy simulation (LES) strategy.\\
 The situation is even more difficult when different turbulent systems are supposed to interact with each other. One of the best-known examples is the temperature field in a turbulent flow, which, depending on the role of buoyancy, may behave as an active or a passive scalar. If very few exact results on the closure problem are known for non interacting turbulent systems, the situation is even worst in the presence of interaction.

 The aim of the present paper is to provide some exact results for a particular class of turbulent systems: the passive-scalar turbulence. This problem is \emph{per se} interesting in connection, e.g., to numerical studies of pollutant concentration in the atmosphere, but it might also give useful suggestions on how to generalize the results to active scalar fields and, hopefully, to hydrodynamic turbulence. This latter point is however, at the present stage, far from being achieved.

 The paper is organized as follow. In \S\ref{sec:pst} we recall the basic features of passive-scalar turbulence. In \S\ref{sec:les} we introduce the large-eddy simulation technique by defining the filtering process, the coarse-grained fields and their subgrid counterparts. The general unclosed character of the equations does not prevent the existence of situations in which exact closures are possible, an example of which is shown in \S\ref{sec:eac}. Starting from \S\ref{sec:kam} we focus our attention on the Kraichnan advection model and, in \S\ref{sec:afe}, we analyse the filtered equation. \S\ref{sec:ac} provides a series of possible closures, emerging with different degrees of approximation or in different contexts. Conclusions and perspectives follow in \S\ref{sec:concl}. Appendices \S\ref{sec:eae} and \S\ref{sec:pdsf} are devoted to provide some exact analytical expressions and to study the limit cases of purely diffusive or smooth flows, respectively.

\section{Passive scalar turbulence} \label{sec:pst}

 The basic equation governing the dynamics of a passive scalar field $\theta(\bm{x},t)$ is the well-known advection-diffusion forced equation
 \begin{equation} \label{basic}
  \partial_t\theta(\bm{x},t)+\bm{v}(\bm{x},t)\cdot\bm{\partial}\theta(\bm{x},t)=\kappa_0\partial^2\theta(\bm{x},t)+f(\bm{x},t)\;.
 \end{equation}
 The advecting velocity field $\bm{v}(\bm{x},t)$ is assumed incompressible: $\bm{\partial}\cdot\bm{v}=0$. Scalar fluctuations are injected into the system at a large scale $L$ by the forcing term $f(\bm{x},t)$, acting as an external source. Scalar dissipation takes place at small scales, of order $\eta$, and is accounted for by the molecular diffusivity $\kappa_0$.\\
 A strong analogy with the Navier--Stokes (NS) turbulence holds: the number of active spatial degrees of freedom \cite{L97} rapidly increases with the P\'eclet number $\textit{Pe}$ (the analogous of the Reynolds number) and scales as $(L/\eta)^d\sim\textit{Pe}^{3d/4}$, where $d$ is the space dimension ($\ge2$). The advantage of dealing with (\ref{basic}) instead of NS consists both in the linearity of the equation and in the scalar character of the unknown field, and finally in the locality of the physical-space description (i.e. the analogous of the pressure field is absent). Nevertheless, the importance of scalar turbulence is underlined by the considerable progress that has been achieved in the last few years in this context \cite{SS00}.

\section{Large-eddy simulation} \label{sec:les}

\subsection{Definition of the filtering process}

 Coarse-grained fields (denoted with a tilde) are obtained from the original, fully-resolved fields through a convolution with a low-pass isotropic filter $P_l$ of characteristic length $l$, lying in the inertial range of scales: $\eta\ll l\ll L$. In particular, we have
 \[\tilde{\theta}(\bm{x},t)\equiv\!\int\!{\rm d}^d\bm{y}\,P_l(\bm{x}-\bm{y})\theta(\bm{y},t)=\!\int\!{\rm d}^d\bm{s}\,P_l(\bm{s})\theta(\bm{x}+\bm{s},t)=(P_l\star\theta)(\bm{x},t)\;,\]
 and similarly for $\tilde{\bm{v}}\equiv P_l\star\bm{v}$ and $\tilde{f}\equiv P_l\star f$. The filtering process thus defines a linear operator which commutates with any partial derivative, because of the structure of the convolution kernel, but which does not factorize when acting on products (the filtered of a product is not the product of the filtered).\\
 Small-scale fluctuations, denoted with a star, are defined as:
 \begin{equation} \label{decom}
  \theta^*\equiv\theta-\tilde{\theta}\;,\qquad\bm{v}^*\equiv\bm{v}-\tilde{\bm{v}}\;,\qquad f^*\equiv f-\tilde{f}\;.
 \end{equation}

\subsection{The problem of closure in the large scale}

 The large-eddy simulation (LES) strategy is carried out by convolving (\ref{basic}) with the filter $P_l$, in order to obtain an equation for the coarse-grained fields:
 \begin{equation} \label{open}
  \partial_t\tilde{\theta}+\widetilde{\bm{v}\cdot\bm{\partial}\theta}=\kappa_0\partial^2\tilde{\theta}+\tilde{f}\;.
 \end{equation}
 Unfortunately, this is not a closed equation in the large scales, as the multiplicative (in the filtering operation) term still involves a product between fully-resolved fields. We thus rearrange it in the form
 \begin{equation} \label{exact}
  \partial_t\tilde{\theta}+\tilde{\bm{v}}\cdot\bm{\partial}\tilde{\theta}=\kappa_0\partial^2\tilde{\theta}+\tilde{f}-\bm{\partial}\cdot\bm{\tau}_{\theta}\;,
 \end{equation}
 where the subgrid scalar flux $\bm{\tau}_{\theta}$ is given by:
 \begin{equation} \label{tau1}
  \bm{\tau}_{\theta}\equiv\widetilde{\bm{v}\theta}-\tilde{\bm{v}}\tilde{\theta}\;.
 \end{equation}
 The aim of LES closures \cite{MK00} is to express $\bm{\partial}\cdot\bm{\tau}_{\theta}$ in terms of coarse-grained fields, in order to get a closed equation describing the large-scale dynamics autonomously. Once this purpose is accomplished, (\ref{exact}) can be integrated numerically (see, e.g., \cite{CLMV01}) on a mesh of spacing $l$ instead of $\eta$, as it would be necessary for the original equation (\ref{basic}). This less-demanding integration means a huge gain in memory and CPU time requirements and represents the essential advantage of the LES strategy.\\
 From a general point of view, the perfect closure (i.e. having the `true' $\bm{\tau}_{\theta}$) is able to correctly represent all the observables built by filtering the `true' field $\theta$. On the contrary, only some observables, and some of them in principle better than others, can be correctly described by any empirical closure. It is worth noticing that, which observable is properly reproduced, can be assessed only \emph{a posteriori} (e.g. by comparing LES predictions against experiments).\\
 Unluckily, no general closed expression for $\bm{\partial}\cdot\bm{\tau}_{\theta}$ in terms of $\tilde{\theta}$ and $\tilde{\bm{v}}$ is available: this is a clear indication of the strong coupling between all scales which is typical in turbulence. A remarkable exception is provided by the case where there is a marked scale separation between velocity and scalar length and time scales. It is then possible to show \cite{BCVV95,M97,MMV05} that the effect of unresolved scales is just the renormalization of the molecular diffusion coefficient $\kappa_0$ to an enhanced eddy diffusivity $\kappa_{\rm eff}$ (generally speaking, an eddy-diffusivity tensorial field). General expressions for the eddy diffusivity as a function of the flow properties do not exist, and in most cases $\kappa_{\rm eff}$ can be determined only numerically.\\
 Here, our aim is to consider the challenging situation where there is no scale separation \cite{MC92} between velocity and scalar and to explore, in such a context, the existence of effective equations for $\tilde{\theta}$.

\subsection{Structure of small-scale contributions}

 An alternative expression of $\bm{\partial}\cdot\bm{\tau}_{\theta}$ can be obtained, for future purpose, plugging in (\ref{tau1}) the decomposition (\ref{decom}). We have:
 \begin{equation} \label{tau2}
  \bm{\partial}\cdot\bm{\tau}_{\theta}=\mathcal{L}+\tilde{\mathcal{C}}+\tilde{\mathcal{R}}\;,
 \end{equation}
 where
 \begin{equation} \label{LCR}
  \left\{\begin{array}{lrl}
   \mathcal{L}\equiv\widetilde{\tilde{\bm{v}}\cdot\bm{\partial}\tilde{\theta}}-\tilde{\bm{v}}\cdot\bm{\partial}\tilde{\theta}&\textrm{(Leonard-like term)}&\\
   \mathcal{C}\equiv\tilde{\bm{v}}\cdot\bm{\partial}\theta^*+\bm{v}^*\cdot\bm{\partial}\tilde{\theta}&\textrm{(cross term)}&\\
   \mathcal{R}\equiv\bm{v}^*\cdot\bm{\partial}\theta^*&\textrm{(Reynolds-like term)}&.
  \end{array}\right.
 \end{equation}

\section{An example of exact closure} \label{sec:eac}

 We now specialize to stochastic velocities and forcings, i.e. to the case in which both $\bm{v}$ and $f$ are fields with assigned statistical properties; we denote with brackets the average over their statistical distribution: it is then clear that this ensemble average commutates with the spatial average represented by the filtering process.\\
 Also in this framework, a renowned example of exact closure exists. Indeed, for times $t$ larger than the largest velocity time scale, the mean field $\langle\theta\rangle$ experiences the cumulative effect of velocity via an eddy-diffusivity coefficient:
 \begin{equation} \label{ktot1}
  \partial_t\langle\theta\rangle=\kappa_{\rm tot}\partial^2\langle\theta\rangle+\langle f\rangle\;.
 \end{equation}
 The expression for $\kappa_{\rm tot}$ (here supposed to be isotropic, as actually is in the presence of isotropic velocity fields) follows from the well-known Taylor formula:
 \begin{equation} \label{kt1}
  \kappa_{\rm tot}=\kappa_0+\frac{1}{2d}\!\int_0^{\infty}\!{\rm d}\tau\,\langle\bm{v}(\tau)\cdot\bm{v}(0)\rangle\;.
 \end{equation}
 Having accounted for the advective term through a linear diffusive one, in this particular case the nonlinearity problem faced while passing from equation (\ref{basic}) to (\ref{open}) does not exist any more and the same eddy-diffusivity equation must then hold also for the averaged coarse-grained field $\langle\tilde{\theta}\rangle$. By virtue of linearity, from (\ref{ktot1}) we have:
 \begin{equation} \label{ktot2}
  \partial_t\langle\tilde{\theta}\rangle=\kappa_{\rm tot}\partial^2\langle\tilde{\theta}\rangle+\langle\tilde{f}\rangle\;.
 \end{equation}
 We now look for an eddy-diffusivity-type closure in the equation for $\tilde{\theta}$ such that, starting from it, (\ref{ktot2}) is recovered. We ask, in other words, that the closure is able to reproduce the averaged, long-time behaviour of $\tilde{\theta}$. The searched equation is
 \begin{equation} \label{keff}
  \partial_t\tilde{\theta}+\tilde{\bm{v}}\cdot\bm{\partial}\tilde{\theta}=\kappa_{\rm eff}\partial^2\tilde{\theta}+\tilde{f}\;,
 \end{equation}
 where $\kappa_{\rm eff}$ has to be determined. We can use Taylor's formula again to obtain
 \begin{equation} \label{kt2}
  \kappa_{\rm tot}=\kappa_{\rm eff}+\frac{1}{2d}\!\int_0^{\infty}\!{\rm d}\tau\,\langle\tilde{\bm{v}}(\tau)\cdot\tilde{\bm{v}}(0)\rangle\;.
 \end{equation}
 A simple comparison between equations (\ref{kt1}) and (\ref{kt2}) yields:
 \begin{equation} \label{ke}
  \kappa_{\rm eff}=\kappa_0+\frac{1}{2d}\!\int_0^{\infty}\!{\rm d}\tau\,\left[\langle\bm{v}(\tau)\cdot\bm{v}(0)\rangle-\langle\tilde{\bm{v}}(\tau)\cdot\tilde{\bm{v}}(0)\rangle\right]\;.
 \end{equation}

\section{Kraichnan advection model} \label{sec:kam}

\subsection{Properties of velocity and forcing}

 We further restrict our attention to the well-known Kraichnan advection model \cite{K68,K94}, in which both velocity and forcing are Gaussian, white-in-time and zero-mean random fields, statistically stationary, homogeneous and isotropic (for an extension to inhomogeneous or anisotropic situations see, e.g., \cite{BP05,MAS05}).\\
 The statistics of $\bm{v}$ is thus fully determined by its second-order increments,
 \[\langle[v_{\mu}(\bm{r},t)-v_{\mu}(\bm{0},0)][v_{\nu}(\bm{r},t)-v_{\nu}(\bm{0},0)]\rangle=2\delta(t)D_{\mu\nu}^{(\bm{v})}(\bm{r})\;,\]
 following an inertial-range spatial power law:
 \begin{equation} \label{Dmunu}
  D_{\mu\nu}^{(\bm{v})}(\bm{r})=D_0r^{\xi}\left[(d+\xi-1)\delta_{\mu\nu}-\xi\frac{r_{\mu}r_{\nu}}{r^2}\right]\;.
 \end{equation}
 The assumption of $\delta$-correlation in time is of course far from reality, but it has the remarkable advantage of leading to closed equations for the equal-time correlation functions $C_n^{(\theta)}\equiv\langle\theta(\bm{x}_1,t)\cdots\theta(\bm{x}_n,t)\rangle$ of any order $n$ (see, e.g., \cite{FGV01}). The parameter $\xi$, lying in the open interval $(0,2)$, governs the roughness of the velocity field, whose H\"{o}lder exponent is $\xi/2$. Due to the lack of temporal memory of the flow, the Kolmogorov value \cite{K41} is $\xi=4/3$; the limit cases $\xi=0$ and $\xi=2$ will be studied in the appendix. The diffusive scale $\eta$, at which diffusive and advective effects are comparable, is defined by the relation:
 \begin{equation} \label{eta}
  \eta\equiv\left(\frac{2\kappa_0}{(d-1)D_0}\right)^{1/\xi}\;.
 \end{equation}
 A convenient choice for $f$ is to assume a step-function form for its two-point correlation $\langle f(\bm{r},t)f(\bm{0},0)\rangle=\delta(t)F_L(r)$, i.e. $F_L(r)=F_0\Theta(L-r)$ (the details of its behaviour around the large scale $L$ are not relevant, but in this case the inertial range has an upper limit $r<L$ instead of $r\ll L$: see, e.g., \cite{FMV98}).

\subsection{Two-point correlation function of the fully-resolved passive scalar}

 In this framework the mean value $\langle\theta\rangle$ shows trivial dynamics, while the equation for the two-point equal-time correlation function $C_2^{(\theta)}$ can be derived analytically from (\ref{basic}) exploiting Furutsu--Novikov--Donsker's (FND) rule for Gaussian integration by parts \cite{F63,N65,D64,F95}:
 \begin{equation} \label{c2t:eq}
  \partial_t C_2^{(\theta)}=D_{\mu\nu}^{(\bm{v})}\partial_{\mu}\partial_{\nu}C_2^{(\theta)}+2\kappa_0\partial^2C_2^{(\theta)}+F_L\;.
 \end{equation}
 The stationary version of (\ref{c2t:eq}), in which the left-hand side vanishes and $C_2^{(\theta)}$ is only a function of the distance $r$ between the two points ($\bm{r}=\bm{x}-\bm{x}'$), can be simply solved by splitting the integration interval and gives \cite{CFKL95}:
 \begin{equation} \label{c2t:sol}
  C_2^{(\theta)}(r)\simeq\left\{\begin{array}{lll}
   c'-k'r^2&\qquad\textrm{for }r\ll\eta&\\
   c-kr^{2-\xi}&\qquad\textrm{for }\eta\ll r<L&\\
   k''r^{-(d+\xi-2)}&\qquad\textrm{for }r>L&,
  \end{array}\right.
 \end{equation}
 where
 \begin{eqnarray*}
  &\displaystyle c=\frac{F_0L^{2-\xi}}{(d-1)(2-\xi)(d+\xi-2)D_0}\;,\qquad c'=c-\frac{\xi F_0\eta^{2-\xi}}{2d(d-1)(2-\xi)D_0}\;,\\
  &\displaystyle k=\frac{F_0}{d(d-1)(2-\xi)D_0}\;,\qquad k'=\frac{2-\xi}{2\eta^{\xi}}k\;,\qquad k''=\frac{(2-\xi)L^d}{d+\xi-2}k\;.
 \end{eqnarray*}
 In the limit of small viscosity, which will be assumed throughout the paper, the merged-point value of the correlation is given by $\langle\theta^2\rangle\simeq c$ and the stationary second-order structure function turns out to be a pure power law in the inertial range:
 \begin{equation} \label{s2t}
  S_2^{(\theta)}(r)\equiv\langle[\theta(\bm{r},t)-\theta(\bm{0},t)]^2\rangle=2\langle\theta^2\rangle-2C_2^{(\theta)}(r)=2kr^{2-\xi}\;.
 \end{equation}
 The exponent $2-\xi$ coincides with the predictions based on dimensional arguments and becomes $2/3$ for $\xi=4/3$, according to the Kolmogorov--Obukhov--Corrsin scaling \cite{MY75}.\\
 The equation for the steady-state dissipation, which we write for future purpose, arises from (\ref{c2t:eq}) evaluated at merged points:
 \begin{equation} \label{diss}
  2\kappa_0\langle(\bm{\partial}\theta)^2\rangle=F_0\;.
 \end{equation}
 Exact solutions, i.e. far from the perturbative limits $\xi\to0$ \cite{GK95}, $\xi\to2$ \cite{SS98} and $d\to\infty$ \cite{CF96}, are not available for higher-order correlations.

\subsection{Two-point correlation function of the coarse-grained passive scalar}

 To provide a benchmark for the various closures, we first evaluate the stationary coarse-grained correlation function by its definition:
 \begin{eqnarray} \label{c2tt:def}
  C_2^{(\tilde{\theta})}(r)&\equiv&\langle\tilde{\theta}(\bm{x},t)\tilde{\theta}(\bm{x}',t)\rangle\nonumber\\
  \!\!&\!\!=\!\!&\!\!\!\int\!{\rm d}^d\bm{y}\!\int\!{\rm d}^d\bm{y}'\,P_l(\bm{x}-\bm{y})P_l(\bm{x}'-\bm{y}')\langle\theta(\bm{y},t)\theta(\bm{y}',t)\rangle\\
  \!\!&\!\!=\!\!&\!\!\!\int\!{\rm d}^d\bm{s}\!\int\!{\rm d}^d\bm{s}'\,P_l(\bm{s})P_l(\bm{s}')C_2^{(\theta)}(|\bm{r}+\bm{s}+\bm{s}'|)\;.\nonumber
 \end{eqnarray}
 Its exact value is reported in the appendix for $d=3$ and a top-hat spherical filter, $P_l(s)=3\Theta(l-s)/4\upi l^3$ (from now on, unless explicitly stated, we will confine ourselves to this situation). Here we only need to express its Taylor's expansion in the parameter $l/r$, with $r$ lying in the inertial range:
 \begin{eqnarray} \label{c2tt:es}
  C_2^{(\tilde{\theta})}(r)=c-kr^{2-\xi}&&\displaystyle\left[1+\frac{1}{5}(2-\xi)(3-\xi)\left(\frac{l}{r}\right)^2\right.\\
  &&\displaystyle\left.+\frac{3}{175}\xi(\xi-1)(2-\xi)(3-\xi)\left(\frac{l}{r}\right)^4+{\rm O}\left(\frac{l}{r}\right)^6\right]\;.\nonumber
 \end{eqnarray}
 Clearly, as the separation $r$ increases and becomes much greater than the filter scale $l$, the unfiltered result is recovered. On the other side, a lower limit for the physical consistence of the latter expansion can be intuitively identified in $r\ge2l$, because for smaller separations the two integration domains in (\ref{c2tt:def}) would partially overlap. Expression (\ref{c2tt:es}) represents, therefore, the best result that can be achieved by means of a closure.\\
 For a 3-D Gaussian filter, $P_l(s)=(2\upi l^2)^{-3/2}{\rm e}^{-s^2/2l^2}$, the coefficient of the $(l/r)^2$ term in (\ref{c2tt:es}) is $(2-\xi)(3-\xi)$. For a 2-D top-hat filter, $P_l(s)=\Theta(l-s)/\upi l^2$, it becomes $(2-\xi)^2/4$.

\subsection{Properties of the filtered velocity and forcing}

 At this stage, it is also convenient to analyse the behaviour of the filtered velocity and forcing. By definition, both $\tilde{\bm{v}}$ and $\tilde{f}$ are Gaussian, white-in-time and zero-mean random fields, statistically stationary, homogeneous and isotropic. The large-scale character of the forcing reflects in the fact that its two-point correlation $\langle\tilde{f}(\bm{r},t)\tilde{f}(\bm{0},0)\rangle=\delta(t)\mathcal{F}_L(r)$ keeps the same spatial step-like form
 \[\mathcal{F}_L(r)=\!\int\!{\rm d}^d\bm{s}\!\int\!{\rm d}^d\bm{s}'\,P_l(\bm{s})P_l(\bm{s}')F_L(|\bm{r}+\bm{s}+\bm{s}'|)=F_L(r)\qquad\textrm{for }r\notin\ell\;,\]
 where the interval $\ell\equiv[L-2l,L+2l]$ becomes negligible for $l\ll L$.\\
 Second-order increments of the coarse-grained velocity,
 \[\langle[\tilde{v}_{\mu}(\bm{r},t)-\tilde{v}_{\mu}(\bm{0},0)][\tilde{v}_{\nu}(\bm{r},t)-\tilde{v}_{\nu}(\bm{0},0)]\rangle=2\delta(t)D_{\mu\nu}^{(\tilde{\bm{v}})}(\bm{r})\;,\]
 are given by
 \begin{eqnarray}
  D_{\mu\nu}^{(\tilde{\bm{v}})}(\bm{r})\!\!&\!\!=\!\!&\!\!\!\int\!{\rm d}^d\bm{s}\!\int\!{\rm d}^d\bm{s}'\,P_l(\bm{s})P_l(\bm{s}')\left[D_{\mu\nu}^{(\bm{v})}(\bm{r}+\bm{s}+\bm{s}')-D_{\mu\nu}^{(\bm{v})}(\bm{s}+\bm{s}')\right]\label{di}\\
  \!\!&\!\!=\!\!&\!\!A(r)\delta_{\mu\nu}+B(r)\frac{r_{\mu}r_{\nu}}{r^2}\;.\label{ab}
 \end{eqnarray}
 The exact expressions of the coefficients $A(r)$ and $B(r)$ are quite cumbersome (see appendix); a much more useful and meaningful quantity is provided by the following power-series expansion in $l/r$:
 \begin{eqnarray} \label{Dmunut}
  D_{\mu\nu}^{(\tilde{\bm{v}})}(\bm{r})\!\!&\!\!=\!\!&\!\!D_0r^{\xi}\left\{\left[(2+\xi)\delta_{\mu\nu}-\xi\frac{r_{\mu}r_{\nu}}{r^2}\right]-\frac{2^{\xi}48\delta_{\mu\nu}}{(4+\xi)(6+\xi)}\left(\frac{l}{r}\right)^{\xi}+{\rm O}\left(\frac{l}{r}\right)^2\right\}\nonumber\\
  \!\!&\!\!=\!\!&\!\!D_{\mu\nu}^{(\bm{v})}(\bm{r})\left[1+{\rm O}\left(\frac{l}{r}\right)^2\right]-\frac{2^{\xi}48D_0l^{\xi}}{(4+\xi)(6+\xi)}\delta_{\mu\nu}\;.
 \end{eqnarray}
 The latter will be extensively used in the following.\\
 Similar results also hold for both the 3-D Gaussian and the 2-D top-hat filter, but with different numerical coefficients.

\section{Analysis of the filtered equation} \label{sec:afe}

 The first step, in order to find closed equations for the large scales, consists in deriving the exact equations for the two-point correlation function of the filtered field. With the same procedure used when passing from (\ref{basic}) to (\ref{c2t:eq}), starting from equation (\ref{exact}) we can write:
 \begin{eqnarray} \label{c2tt:eq}
  &\partial_t\langle\tilde{\theta}(\bm{x},t)\tilde{\theta}(\bm{x}',t)\rangle+2\langle\tilde{\theta}(\bm{x},t)(\tilde{\bm{v}}\cdot\bm{\partial}\tilde{\theta})(\bm{x}',t)\rangle=\nonumber\\
  &=2\kappa_0\partial^2\langle\tilde{\theta}(\bm{x},t)\tilde{\theta}(\bm{x}',t)\rangle+\mathcal{F}_L(|\bm{x}-\bm{x}'|)-2\langle\tilde{\theta}(\bm{x},t)\bm{\partial}\cdot\bm{\tau}_{\theta}(\bm{x}',t)\rangle\;.
 \end{eqnarray}
 This is the starting point for our systematic procedure to construct closure approximations. The second term on the left-hand side depends on $\tilde{\bm{v}}$ and $\tilde{\theta}$ only, but we cannot transform it (yet) into a contribution with the structure of the first term on the right-hand side of (\ref{c2t:eq}) because, at this stage, we are not able to apply FND's rule on it, as we do not know the functional derivative of $\tilde{\theta}$ with respect to $\tilde{\bm{v}}$ explicitly. The last term on the right-hand side of (\ref{c2tt:eq}), which is not expressed as a function of coarse-grained fields, is the `disturbing' quantity: our purpose is therefore to find its perturbative expansion in $l/r$.\\
 It is not difficult (although quite lengthy) to prove that:
 \begin{equation} \label{tau}
  \langle\tilde{\theta}(\bm{x},t)\bm{\partial}\cdot\bm{\tau}_{\theta}(\bm{x}',t)\rangle=\frac{2^{\xi}4(3-\xi)F_0}{(4+\xi)(6+\xi)}\left(\frac{l}{r}\right)^{\xi}-\frac{\xi^2F_0}{30}\left(\frac{l}{r}\right)^2+{\rm O}\left(\frac{l}{r}\right)^{2+\xi}\;.
 \end{equation}
 More specifically, exploiting decomposition (\ref{tau2}) and definitions (\ref{LCR}), we can show that (\ref{tau}) consists of:
 \begin{equation} \label{tauLCR}
  \left\{\begin{array}{lll}
   \langle\mathcal{L}(\bm{x},t)\tilde{\theta}(\bm{x}',t)\rangle=\!\!&\!\!\displaystyle{\rm O}\left(\frac{l}{r}\right)^{2+\xi}&\\
   \langle\mathcal{C}(\bm{x},t)\tilde{\theta}(\bm{x}',t)\rangle=\!\!&\!\!\displaystyle-\frac{(3-\xi)(\xi^2+10\xi+24-2^{\xi}24)F_0}{3(4+\xi)(6+\xi)}\left(\frac{l}{r}\right)^{\xi}\\
   &\!\!\displaystyle\quad-\frac{\xi^2F_0}{30}\left(\frac{l}{r}\right)^2+{\rm O}\left(\frac{l}{r}\right)^{2+\xi}&\\
   \langle\mathcal{R}(\bm{x},t)\tilde{\theta}(\bm{x}',t)\rangle=\!\!&\!\!\displaystyle\frac{(3-\xi)(\xi^2+10\xi+24-2^{\xi}12)F_0}{3(4+\xi)(6+\xi)}\left(\frac{l}{r}\right)^{\xi}+{\rm O}\left(\frac{l}{r}\right)^{2+\xi}&.
  \end{array}\right.
 \end{equation}
 The first line of (\ref{tauLCR}) tells us that the Leonard-type term does not contribute, at the lowest two orders, to the equation for the two-point correlation of the coarse-grained scalar. Since our closures are derived from this equation up to the second order, it follows that the Leonard-type term will not contribute to small-scale parameterizations. This fact is not a consequence of the Kraichnan advection model but rather seems to hold for general advection models. For standard closure models based on single-point quantities, the contribution from the Leonard stress in the parameterizations is, generally speaking, non-zero.\\
 Moreover, it is easy to see that the sum of the last two equations in (\ref{tauLCR}) exactly coincides with (\ref{tau}), in spite of the further convolution required in (\ref{tau2}) on the cross and Reynolds-like terms. This is in accordance with the following result, that can be obtained after simple algebra:
 \[\langle\mathcal{C}(\bm{x},t)\theta(\bm{x}',t)\rangle+{\rm O}\left(\frac{l}{r}\right)^{2+\xi}=\langle\mathcal{C}(\bm{x},t)\tilde{\theta}(\bm{x}',t)\rangle=\langle\tilde{\mathcal{C}}(\bm{x},t)\tilde{\theta}(\bm{x}',t)\rangle+{\rm O}\left(\frac{l}{r}\right)^{2+\xi}\;.\]
 The same result holds replacing $\mathcal{C}$ with $\mathcal{R}$ and has been obtained exploiting a further identity, holding for any couple of fields $\bm{f}_1$ and $\bm{f}_2$ in the presence of statistical homogeneity: $\langle\tilde{\bm{f}_1}(\bm{r},t)\bm{f}_2(\bm{0},0)\rangle=\langle\bm{f}_1(\bm{r},t)\tilde{\bm{f}_2}(\bm{0},0)\rangle$.\\
 It is also interesting to study when the cross term and the Reynolds-like one dominate over each other. At the lowest order, $(l/r)^{\xi}$, we should thus compare the absolute values of the two numerical coefficients (functions of $\xi$) appearing on the second and on the last line of (\ref{tauLCR}): the result is that the former prevails for $\xi>\xi_0$, with $\xi_0\simeq0.92$, and the latter is dominant when $\xi$ does not exceed this critical value $\xi_0$. 
 Furthermore, one can easily prove that the two terms composing $\mathcal{C}$ (\ref{LCR}) give the same contribution to the parameterization, each being half of the numerical coefficient appearing on the second line of (\ref{tauLCR}).\\
 Once again, expressions with the same structure of (\ref{tau}) and (\ref{tauLCR}) also hold for the 3-D Gaussian filter, with different numerical coefficients. On the contrary, for the 2-D top-hat filter, it turns out that the second-order coefficient in (\ref{tau}) vanishes.

\section{Examples of analytical closures} \label{sec:ac}

\subsection{Importance of small-scale contributions}

 Let us now focus on consequences of relation (\ref{tau}): it is immediate to realize that neglecting small-scales effects completely, i.e. assuming $\bm{\partial}\cdot\bm{\tau}_{\theta}=0$, makes equation (\ref{c2tt:eq}) unbalanced at order $(l/r)^{\xi}$. In other words, if one assumes a closure of the kind
 \[\partial_t\tilde{\theta}+\tilde{\bm{v}}\cdot\bm{\partial}\tilde{\theta}=\kappa_0\partial^2\tilde{\theta}+\tilde{f}\;,\]
 starting from it one would obtain a non-analytical expansion for the coarse-grained scalar correlation, $C_2^{(\tilde{\theta})}(r)=c-kr^{2-\xi}[1+{\rm O}(l/r)^{\xi}]$, clearly in contrast with the exact result (\ref{c2tt:es}).

\subsection{Constant-eddy-diffusivity closure} \label{ced}

 The first issue thus consists in finding a way to take order $(l/r)^{\xi}$ into account properly. For this purpose it is sufficient to notice that the diffusive contribution in (\ref{c2tt:eq}), or equivalently (at this order of approximation) in (\ref{c2t:eq}), turns out to be proportional to $r^{-\xi}$, precisely
 \begin{equation} \label{k0}
  \kappa_0\partial^2C_2^{(\theta)}=-\frac{(3-\xi)F_0\kappa_0}{6D_0r^{\xi}}\;.
 \end{equation}
 It is then clear that an effective-diffusivity term, like the one proposed in (\ref{keff}), must be able to correctly reproduce the lowest-order contribution of (\ref{tau}). More specifically, writing $\kappa_{\rm eff}=\kappa_0+\kappa_1$, the equation
 \begin{equation} \label{CED}
  \partial_t\tilde{\theta}+\tilde{\bm{v}}\cdot\bm{\partial}\tilde{\theta}=\kappa_{\rm eff}\partial^2\tilde{\theta}+\tilde{f}
 \end{equation}
 is balanced at order $(l/r)^{\xi}$ if
 \begin{equation} \label{k1}
  \kappa_1=\frac{2^{\xi}24}{(4+\xi)(6+\xi)}D_0l^{\xi}
 \end{equation}
 (to prove this, it is sufficient to replace $\kappa_0$ with $\kappa_1$ in (\ref{k0}) and to compare the result with (\ref{tau})). It is worth noticing that expression (\ref{k1}) also follows from the integral in (\ref{ke}), which substantially amounts to compute the difference between total and large-scale kinetic energies in the presence of $\delta$-correlated flows. This is in accordance with equation (\ref{Dmunut}), from which we deduce
 \begin{equation} \label{kd}
  \kappa_1\delta_{\mu\nu}=\lim_{r\to\infty}\frac{D_{\mu\nu}^{(\bm{v})}(\bm{r})-D_{\mu\nu}^{(\tilde{\bm{v}})}(\bm{r})}{2}\;.
 \end{equation}
 Two more remarks about this kind of closure, which we will call `constant eddy diffusivity' (CED), emerge from the consideration that the fraction (function of $\xi$) in (\ref{k1}) always stands between 1 and 2. First, reminding that from (\ref{eta}) we have $\kappa_0=D_0\eta^{\xi}$, a remarkable increase in the transport coefficient is found ($l\gg\eta\Rightarrow\kappa_{\rm eff}\simeq\kappa_1\gg\kappa_0$): this is a typical effect in turbulence. Second, an effective dissipative scale comparable to the filtering length $l$ arises for $\tilde{\theta}$: the analogy is evident between the roles played by $l$ for $\tilde{\theta}$ and by the molecular dissipative scale $\eta$ for the original field $\theta$.\\
 The equation for the coarse-grained scalar correlation arising from (\ref{CED}) has exactly the same structure of the fully-resolved corresponding (\ref{c2t:eq}):
 \begin{equation} \label{CED:eq}
  \partial_t C_2^{(\tilde{\theta})}=D_{\mu\nu}^{(\tilde{\bm{v}})}\partial_{\mu}\partial_{\nu}C_2^{(\tilde{\theta})}+2\kappa_{\rm eff}\partial^2C_2^{(\tilde{\theta})}+\mathcal{F}_L\;.
 \end{equation}
 The stationary solution of (\ref{CED:eq}) in the inertial range is:
 \begin{equation} \label{CED:sol}
  C_2^{(\tilde{\theta})}(r)=c-kr^{2-\xi}\left[1+\frac{1}{5}(2-\xi)(3+\xi)\left(\frac{l}{r}\right)^2+{\rm O}\left(\frac{l}{r}\right)^4\right]\;.
 \end{equation}
 A comparison with the exact result (\ref{c2tt:es}) shows that the CED closure is able to capture the correct order of the deviation from the fully-resolved scalar correlation (\ref{c2t:sol}), i.e. $(l/r)^2$, but with a wrong coefficient. It can be proved that the maximum error takes place at $\xi=1$.\\
 From (\ref{CED:eq}) we derive the following equation for the steady-state dissipation:
 \begin{equation} \label{dissip}
  2\kappa_{\rm eff}\langle(\bm{\partial}\tilde{\theta})^2\rangle=F_0\;.
 \end{equation}
 Owing to the fact that $\kappa_{\rm eff}\gg\kappa_0$, the comparison of (\ref{diss}) with (\ref{dissip}) tells us that the average of the square gradient of the scalar is much smaller for the filtered field than for the original one, as one would expect.\\
 For the 3-D Gaussian filter or the 2-D top-hat filter, expression (\ref{k1}) becomes
 $$\kappa_1=\frac{2^{1+\xi}}{3\sqrt{\pi}}(3+\xi)\Gamma\left(\frac{3+\xi}{2}\right)D_0l^{\xi}$$
 or
 $$\kappa_1=\frac{2+\xi}{2}\frac{\Gamma(2+\xi)}{\Gamma(2+\xi/2)\Gamma(3+\xi/2)}D_0l^{\xi}$$
 respectively, $\Gamma$ being Euler's function. Consequently, the ratio $\kappa_1/D_0l^{\xi}$ stands between $1$ and $10$ in the former case and between $1/2$ and $1$ in the latter. Plugging these values of $\kappa_1$ in equation (\ref{CED}) or (\ref{CED:eq}) and computing the coarse-grained correlation, for the 3-D Gaussian filter we obtain a result similar to (\ref{CED:sol}), without the $1/5$ factor in the $(l/r)^2$ term. However, for the 2-D top-hat filter, the second-order coefficient is $(2-\xi)^2/4$ and exactly coincides with its `true' value, i.e. we get the remarkable result that the error in the approximation is automatically pushed at higher orders.

\subsection{Improved closure} \label{ded}

 Our aim is now to improve the constant-eddy-diffusivity closure, exact at order $(l/r)^{\xi}$, by introducing a new closure which is accurate up to order $(l/r)^2$ \cite{MACFM03}. At this stage, we have no hint of how to implement this closure, differently from what happens with the intuitive emergence of the eddy diffusivity; a trivial Taylor expansion on the turbulent fields would actually prove wrong.\\
 However, we know that we have to reproduce a term proportional to $l^2$ in the equation for the coarse-grained scalar correlation: it is then reasonable to add, on the right-hand side of the equation for $\tilde{\theta}$, a new contribution proportional to some power of $l$. The minimal guess could be represented by the addition of a term linear in $l$, which would be able to generate a quadratic correction in the equation for the correlation when applying FND's rule. This guess is however ruled out by symmetry considerations, because we would need to introduce some additional field (with the dimensions of the vorticity) which cannot appear in the equation for $\tilde{\theta}$ for parity conservation.\\
 The next possibility is thus to add a term quadratic in $l$, in which the coarse-grained fields must appear in the tensorial form $\tilde{\bm{v}}\bm{\partial}\tilde{\theta}$. Dimensional considerations then require the presence of a square length at denominator, but no scales other than the filter width can appear, because we are dealing with a single-point equation and neither $L$ nor $\eta$ are relevant. A second derivative is thus required. If one completely neglects higher orders in $l$ (it is important to underline that this ansatz is not trivial at all, because they would give rise to spurious contributions at lower orders), then the searched equation must have the following form:
 \begin{equation} \label{DED1}
  \partial_t\tilde{\theta}+\tilde{\bm{v}}\cdot\bm{\partial}\tilde{\theta}=\kappa_{\rm eff}\partial^2\tilde{\theta}+\tilde{f}+l^2\left(\alpha\partial^2\tilde{\bm{v}}\cdot\bm{\partial}\tilde{\theta}+\beta\bm{\partial}\tilde{\bm{v}}:\bm{\partial}\bm{\partial}\tilde{\theta}+\gamma\tilde{\bm{v}}\cdot\bm{\partial}\partial^2\tilde{\theta}\right)\;.
 \end{equation}
 The coefficients $\alpha$, $\beta$ and $\gamma$ can be uniquely found by imposing the correct description of order $(l/r)^2$ and, at the same time, the vanishing of any spurious modification to order $(l/r)^{\xi}$, which has already been captured by the CED closure. In other words, we ask that the value of $\kappa_{\rm eff}$ previously found remains unchanged in (\ref{DED1}) (alternative conditions will be shown later in this section). With these hypotheses, we have $\alpha=0$ and $\beta=\gamma=-\xi/15$, so that (\ref{DED1}) can be written as:
 \begin{equation} \label{DED2}
  \partial_t\tilde{\theta}+\tilde{\bm{v}}\cdot\bm{\partial}\tilde{\theta}=\kappa_{\rm eff}\partial^2\tilde{\theta}+\tilde{f}-\frac{\xi}{15}l^2\partial_{\mu}\partial_{\nu}(\tilde{v}_{\mu}\partial_{\nu}\tilde{\theta})
 \end{equation}
 (for a comparison with the corresponding nonlinear closure in NS turbulence see, e.g., \cite{BO98,KM01}). Note, however, that (\ref{DED2}) does not coincide with equation (3.16) reported in \cite{MACFM03}. The reason of such a discrepancy has been found in a wrong identification done in \cite{MACFM03} to infer the closure (see, e.g., \cite{MACM04}).\\
 It is worth noticing that this closure, which has been obtained through Eulerian considerations, has no Lagrangian counterpart, differently from CED. Indeed, an expansion in the spirit of Kramers--Moyal would yield a term with the same structure of the one we have introduced on the right-hand side of (\ref{DED1}), but it turns out that our triplet of coefficients $\alpha$, $\beta$, $\gamma$ appearing in (\ref{DED2}) does not satisfy the constraints imposed by Pawula's theorem \cite{R89,G85}. This fact is also related to the breaking of Galilean invariance in (\ref{DED2}), which has been obtained in the frame of reference where the velocity is zero-mean and is not exportable. Moreover, the universality of $\kappa_{\rm eff}$ is no more present in the triplet, whose value has been determined exploiting the explicit solution for $C_2^{(\theta)}$.\\
 The equation for the two-point equal-time correlation function arising from (\ref{DED2}) has now a different structure:
 \begin{eqnarray} \label{DED:eq}
  &\displaystyle\partial_t C_2^{(\tilde{\theta})}=D_{\mu\nu}^{(\tilde{\bm{v}})}\partial_{\mu}\partial_{\nu}C_2^{(\tilde{\theta})}+2\kappa_{\rm eff}\partial^2C_2^{(\tilde{\theta})}+\mathcal{F}_L+\frac{2\xi}{15}l^2\partial_{\lambda}\left[D_{\mu\nu}^{(\tilde{\bm{v}})}\partial_{\lambda}\partial_{\mu}\partial_{\nu}C_2^{(\tilde{\theta})}\right]+\nonumber\\
  &\displaystyle+\frac{\xi}{225}l^4\left\{\partial_{\gamma}\partial_{\lambda}\left[D_{\mu\nu}^{(\tilde{\bm{v}})}\partial_{\gamma}\partial_{\lambda}\partial_{\mu}\partial_{\nu}C_2^{(\tilde{\theta})}\right]+V_{\gamma\lambda\mu\nu}\partial_{\gamma}\partial_{\lambda}\partial_{\mu}\partial_{\nu}C_2^{(\tilde{\theta})}\right\}\;,
 \end{eqnarray}
 where $V_{\gamma\lambda\mu\nu}$ satisfies $\langle\tilde{v}_{\mu}(\bm{0},t)\partial_{\gamma}\partial_{\lambda}\tilde{v}_{\nu}(\bm{0},0)\rangle=\delta(t)V_{\gamma\lambda\mu\nu}$.\\
 The stationary solution of (\ref{DED:eq}) in the inertial range is:
 \begin{equation} \label{DED:sol}
  C_2^{(\tilde{\theta})}(r)=c-kr^{2-\xi}\left[1+\frac{1}{5}(2-\xi)(3-\xi)\left(\frac{l}{r}\right)^2+{\rm O}\left(\frac{l}{r}\right)^{2+\xi}\right]\;.
 \end{equation}
 Comparing (\ref{DED:sol}) with (\ref{c2tt:es}) we conclude that this new kind of closure is able to reproduce the exact structure of the coarse-grained scalar correlation up to the second order with the correct value of the coefficient, differently from what happens in (\ref{CED:sol}). We also notice that the error is now ${\rm O}(l/r)^{2+\xi}$ instead of ${\rm O}(l/r)^4$: in order to balance also this contribution in the proper way, one would have to add other terms on the right-hand side of (\ref{DED2}), paying attention to take unmodified the lower orders which have already been captured.\\
 A last comment on (\ref{DED2}) is worthwhile. It turns out that the same dissipation equation (\ref{dissip}) generated from CED still holds. This is in accordance with the flux-like structure of the last term on the right-hand side of (\ref{DED2}), which gives no contribution to the equation for the dissipation. Exploiting homogeneity (H) and incompressibility (I), we have indeed:
 \[\begin{array}{cccccl}
  \langle\tilde{\theta}\partial_{\mu}\partial_{\nu}(\tilde{v}_{\mu}\partial_{\nu}\tilde{\theta})\rangle&\stackrel{\rm I}{=}&\langle\tilde{\theta}\partial_{\nu}(\tilde{v}_{\mu}\partial_{\mu}\partial_{\nu}\tilde{\theta})\rangle&\stackrel{\rm H}{=}&-\langle(\partial_{\nu}\tilde{\theta})\tilde{v}_{\mu}\partial_{\mu}\partial_{\nu}\tilde{\theta}\rangle&\\
  &\stackrel{\rm I}{=}&-\langle(\partial_{\nu}\tilde{\theta})\partial_{\mu}(\tilde{v}_{\mu}\partial_{\nu}\tilde{\theta})\rangle&\stackrel{\rm H}{=}&\langle(\partial_{\mu}\partial_{\nu}\tilde{\theta})\tilde{v}_{\mu}\partial_{\nu}\tilde{\theta}\rangle&.
 \end{array}\]
 Comparing the third and the last member of this chain of equalities, we deduce that all terms must vanish; in particular, also the first one is thus zero, which proves our assertion.\\
 For the 3-D Gaussian filter, the correct values of the triplet are $\alpha=0$ and $\beta=\gamma=-\xi/3$. The closed equation for $\tilde{\theta}$ has thus the same structure of (\ref{DED2}) and leads to a coarse-grained correlation expressed by (\ref{DED:sol}) but without the $1/5$ factor near $(l/r)^2$, which is exact up to the second-order. The same accuracy is obtained for the 2-D top-hat filter with $\alpha=\beta=\gamma=0$, coherently with what stated at the end of \S\ref{sec:afe} and of \S\ref{ced}.\\
 Before proceeding, we would like to stress the conditions we specified when (uniquely) fixing the values of $\alpha$, $\beta$ and $\gamma$ in equation (\ref{DED1}) in order to obtain (\ref{DED2}): 1) correct description of ${\rm O}(l/r)^2$, 2) complete neglection of terms containing powers of $l$ higher than $2$, and 3) absence of any modification of ${\rm O}(l/r)^{\xi}$ i.e. of $\kappa_{\rm eff}$.

\subsection{An alternative closure}

 We are now going to show an example of what happens if the possibility of taking filters of (slightly) different length between scalar and velocity is allowed. In particular, if we define a new coarse-grained velocity as $\hat{\bm{v}}\equiv P_{al}*\bm{v}$ (with $a=\sqrt{(3-\xi)/(3+\xi)}$, i.e. slightly less than unity), then we assert that an equation with the same structure as in (\ref{CED}),
 \[\partial_t\tilde{\theta}+\hat{\bm{v}}\cdot\bm{\partial}\tilde{\theta}=\kappa_{\rm eff}\partial^2\tilde{\theta}+\tilde{f}\;,\]
 is able to reproduce for the correlation the same degree of approximation of (\ref{DED:sol}), with an error pushed again at ${\rm O}(l/r)^4$ and without any need of introducing additive terms like on the right-hand side of (\ref{DED1}). The same expression (\ref{ke}) for the eddy diffusivity still holds, with $\tilde{\bm{v}}$ replaced by $\hat{\bm{v}}$; this amounts to perform the same replacing in expression (\ref{kd}) for $\kappa_1=\kappa_{\rm eff}-\kappa_0$, whose numerical value is now $a^{\xi}$ times the value given by (\ref{k1}), in accordance with the substitution $l\mapsto al$.

\subsection{Absence of eddy diffusivity}

 We now present some other possible closures, which emerge if one pursues aims different from those specified previously.\\
 From another point of view, the condition 3) fixed at the end of \S\ref{ded} could be modified into: 3bis) modification of ${\rm O}(l/r)^{\xi}$ such that $\kappa_{\rm eff}$ is restored to its original value $\kappa_0$. In other words, we would now like to pass from equation (\ref{exact}) directly to
 \begin{equation} \label{DED3}
  \partial_t\tilde{\theta}+\tilde{\bm{v}}\cdot\bm{\partial}\tilde{\theta}=\kappa_0\partial^2\tilde{\theta}+\tilde{f}+l^2\left(\alpha'\partial^2\tilde{\bm{v}}\cdot\bm{\partial}\tilde{\theta}+\beta'\bm{\partial}\tilde{\bm{v}}:\bm{\partial}\bm{\partial}\tilde{\theta}+\gamma'\tilde{\bm{v}}\cdot\bm{\partial}\partial^2\tilde{\theta}\right)\;,
 \end{equation}
 imposing that the contribution in parentheses on the right-hand side of (\ref{DED3}) is able to correctly balance both ${\rm O}(l/r)^{\xi}$ and ${\rm O}(l/r)^2$. This goal can be accomplished for
 \begin{eqnarray*}
  &\displaystyle\alpha'=\frac{\xi+2}{\xi(3+\xi)(6+\xi)}\;,\qquad\beta'=\frac{-\xi^4-9\xi^3-18\xi^2+120\xi+240}{15\xi(3+\xi)(6+\xi)}\;,\\
  &\displaystyle\gamma'=\frac{-\xi^4-9\xi^3-18\xi^2+60\xi+120}{15\xi(3+\xi)(6+\xi)}\;.
 \end{eqnarray*}
 It is worth noticing that, in this way, the term proportional to $l^2$ in (\ref{DED3}) has no more an overall flux-like structure, differently from what happened in (\ref{DED2}). Consequently, it gives rise to a non-zero contribution in the equation for the dissipation.\\
 Conditions 3) and 3bis) represent two completely different points of view: in the first one the diffusive term captures ${\rm O}(l/r)^{\xi}$ completely, in the second one it gives no contribution. Of course, there exists an infinite range of intermediate possibilities if one considers both the renormalized diffusivity and the triplet as unknowns.

\subsection{Absence of second-order correction}

 Suppose now to be interested to measure `pure' inertial-range scaling behaviour by means of a LES strategy. More specifically, referring to (\ref{c2t:sol}), our aim here is to reproduce the asymptotic behaviour $C_2^{(\theta)}(r)\sim r^{2-\xi}$ even in the presence of finite-size effects, i.e. induced by the filter cut-off $l$, which would imply the additive corrections in powers of $l/r$ shown in (\ref{c2tt:es}). In plain words, we renounce to describe the actual coarse-grained correlation function (\ref{c2tt:es}), with the aim of isolating the scaling behaviour which would be observed at higher-Reynolds-number (and thus higher-resolution) numerical simulations.\\
 Assuming the conditions 2) and 3) mentioned at the end of \S\ref{ded}, this amounts to modify condition 1) into: 1bis) absence of second-order correction in the coarse-grained correlation. This aim is accomplished simply assuming $\alpha=0$ and $\beta=\gamma=-(3+\xi)/30$.\\
 It can be shown indeed that, starting from the equation
 \[\partial_t\tilde{\theta}+\tilde{\bm{v}}\cdot\bm{\partial}\tilde{\theta}=\kappa_{\rm eff}\partial^2\tilde{\theta}+\tilde{f}-\frac{3+\xi}{30}l^2\partial_{\mu}\partial_{\nu}(\tilde{v}_{\mu}\partial_{\nu}\tilde{\theta})\;,\]
 the coarse-grained correlation turns out to be:
 \[C_2^{(\tilde{\theta})}(r)=c-kr^{2-\xi}\left[1+{\rm O}\left(\frac{l}{r}\right)^{2+\xi}\right]\;.\]
 The deviation from the fully-resolved correlation is pushed at a higher order: we have thus obtained a better recovery of the original field in its two-point correlation function.

\section{Conclusions} \label{sec:concl}

 The main aim of the present paper was to provide a systematic procedure to derive closed equations for a coarse-grained passive scalar field. Such equations have been shown in \S\ref{sec:ac} and, starting from each of them, we computed the corresponding two-point equal-time scalar correlation function to which they give rise, in the stationary state of the Kraichnan advection model. Specifically, in \S\ref{ced} we performed an analytical derivation of the `standard' constant-eddy-diffusivity parameterization, which also arises in several other contexts but which is usually obtained empirically. Our procedure allowed us not only to quantify the error intrinsic in the parameterization, by constantly keeping it under control, but also to refine this closure by introducing a new term (\S\ref{ded}) in the equation for $\tilde{\theta}$, thus obtaining a dynamical parameterization which can be seen as the passive-scalar counterpart of the nonlinear closure in NS turbulence. In this way, we were able to correctly describe both the asymptotic behaviour and the scales of motion $r$ in the vicinity of the filter cut-off $l$ (a limit $r\ge2l$ can roughly be identified for the validity of this description), as this closure captures the ${\rm O}(l/r)^2$ correction and pushes the error at higher orders. A natural question then arises: is it possible to generalize the aforementioned procedure, in order to correctly describe also higher orders? This problem is twofold, according to the meaning of the word `order'.\\
 First, always focusing on the two-point correlation, one would like to be able to consider separations $r$ closer and closer to the filtering length $l$ (or, equivalently, to use a larger, numerically less-demanding, filtering length, to describe the same separation): this could be accomplished by balancing equation (\ref{c2tt:eq}) also at some further orders, e.g. ${\rm O}(l/r)^{2+\xi}$ and ${\rm O}(l/r)^4$. This task looks feasible by adding another term on the right-hand side of (\ref{DED2}), proportional to $l^4$ and whose tensorial structure could be determined by symmetry considerations. However, finding the exact numerical coefficients (the counterpart of the triplet $\alpha$, $\beta$, $\gamma$) would imply cumbersome calculations, and --- in view of what stated before equation (\ref{DED1}) and in condition 2) at the end of \S\ref{ded} --- spurious effects at lower orders may appear.\\
 A second, and probably much more meaningful, generalization of the present work would be represented by the extension to higher-order (always equal-time) correlation functions, e.g. $C_4$. Our analysis has been carried out for the second-order (i.e. two-point) correlation function of the scalar field: there are two reasons for this choice. First of all, the second-order correlation function is the Fourier transform of the spectrum of scalar variance, a statistical indicator widely used to characterize most of the statistical properties of scalar turbulence. Moreover, for the Kraichnan model only the second-order correlation function has a simple, closed analytical expression; for higher-order correlations only perturbative expressions (for example in the limit of small $\xi$) are available \cite{GK95,FGV01}. This latter fact suggests that, while a constant-eddy-diffusivity may be found also for them, the exportability of our refined closure seems ruled out, in view of the considerations reported before equation (\ref{DED:eq}). Furthermore, even if the functional form of the closure was preserved, a modification of the effective coefficients would mean that strong small-scale fluctuations --- associated to higher-order correlation functions --- must be described by parameters different from the ones used for less intense fluctuations. That would question the applicability of closure models to the description of the statistics of turbulent fields as temperature or concentration, which are characterized by a wide range of fluctuation intensities.\\
 Let us also mention another possible extension, in the presence of inhomogeneities or anisotropies. Starting from some known results about the two-point equal-time correlation function in these nonideal situations \cite{BP05,MAS05}, it would be of interest to derive the corresponding LES closures from first principles.

 The two above-mentioned closures, (\ref{CED}) and (\ref{DED2}), are `universal' in the sense that they were derived from first principles. However, in the remaining subsections of \S\ref{ded}, we showed that some other closures may emerge. A slight diversification in the filtering length between velocity and scalar allows to capture the ${\rm O}(l/r)^2$ correction even imposing a constant-eddy-diffusivity parameterization. On the other hand, it is possible to obtain the same accuracy doing without any renormalized diffusion coefficient. Finally, we showed that our refined model can be applied to other quantities than the coarse-grained correlation function: e.g., with a simple change in the numerical coefficients, one can reproduce the original, fully-resolved correlation, i.e. the `pure' inertial-range scaling, even in the presence of finite-size effects.

 A last question that naturally arises is whether these results are relevant for realistic advection models, e.g. relaxing the assumption of $\delta$-correlation in time. Of course, the value of the effective diffusivity and of the triplet of numerical coefficients that appear in these closures can be analytically computed only in the Kraichnan model. However, we believe that the form of the parameterization can be exported without modifications to real situations as well. Clearly, in this case, such free parameters have to be determined \emph{a posteriori} by some empirical procedure. The validity of this approach has already been checked by direct numerical simulations \cite{MACM04}, which confirm such a preview even with velocity fields that are real solutions of the NS equation.

\section*{Acknowledgements}

 AM and MMA have been partially supported by COFIN 2005 under the project n.~2005027808 and by CINFAI consortium.  Part of this work has been done within the 2006 CNR Short-Term Mobility programme (AM).

\appendix

\section{Exact analytical expressions} \label{sec:eae}

 In this section we show the exact analytical expressions of some quantities, of which we have only reported perturbative expressions throughout the paper.

\subsection{Coarse-grained passive scalar}

 The correlation of the coarse-grained passive scalar mentioned in (\ref{c2tt:es}) is:
 \begin{eqnarray*}
  C_2^{(\tilde{\theta})}(r)\!\!&\!\!=\!\!&\!\!c-\frac{3F_0}{8(2-\xi)(4-\xi)(5-\xi)(6-\xi)D_0l^6}\left\{-\frac{l}{7-\xi}\left[R_+^{7-\xi}-R|R_-|^{7-\xi}\right]\right.\\
  &\!\!&\!\!-\frac{l^2}{(7-\xi)r}\left[R_+^{7-\xi}+R|R_-|^{7-\xi}-2r^{7-\xi}\right]+\frac{l}{(8-\xi)r}\left[R_+^{8-\xi}-|R_-|^{8-\xi}\right]\!\\
  &\!\!&\!\!+\frac{1}{(7-\xi)(8-\xi)}\left[R_+^{8-\xi}+|R_-|^{8-\xi}-2r^{8-\xi}\right]+\frac{l}{(7-\xi)(8-\xi)r}\left[R_+^{8-\xi}\right.\\
  &\!\!&\!\!\left.\left.-|R_-|^{8-\xi}\right]-\frac{1}{(7-\xi)(9-\xi)r}\left[R_+^{9-\xi}+R|R_-|^{9-\xi}-2r^{9-\xi}\right]\right\}\;,
 \end{eqnarray*}
 where $R_{\pm}=r\pm2l$ and $R=\textrm{sgn}(R_-)$. The fuse-point value is given by
 \[\langle\tilde{\theta}^2\rangle=c-\frac{48F_0l^{2-\xi}}{2^{\xi}(2-\xi)(4-\xi)(5-\xi)(6-\xi)D_0}\;,\]
 i.e. $\langle\tilde{\theta}^2\rangle<\langle\theta^2\rangle=c$.\\
 Keeping (\ref{s2t}) and (\ref{c2tt:es}) into account, this leads to the following expression for the two-point coarse-grained scalar structure function in the inertial range:
 \begin{eqnarray} \label{s2tt}
  S_2^{(\tilde{\theta})}(r)&\equiv&\langle[\tilde{\theta}(\bm{r},t)-\tilde{\theta}(\bm{0},t)]^2\rangle=2\langle\tilde{\theta}^2\rangle-2C_2^{(\tilde{\theta})}(r)=\nonumber\\
  \!\!&\!\!=\!\!&\!\!2\left[c-\frac{48F_0l^{2-\xi}}{2^{\xi}(2-\xi)(4-\xi)(5-\xi)(6-\xi)D_0}\right]\nonumber\\
  &&\quad-2\left\{c-kr^{2-\xi}\left[1+{\rm O}\left(\frac{l}{r}\right)^2\right]\right\}\nonumber\\
  \!\!&\!\!=\!\!&\!\!S_2^{(\theta)}(r)\left[1+{\rm O}\left(\frac{l}{r}\right)^{2-\xi}\right]\;.
 \end{eqnarray}
 Expression (\ref{s2tt}) shows that spurious corrections ${\rm O}(l/r)^{2-\xi}$, which are absent in the correlation, appear in the structure function when filtering. In other words, it is not true that the coarse-grained structure function is obtained by performing a double convolution on the fully-resolved corresponding quantity (this result is in accordance with the integral expression of the coarse-grained velocity structure function in (\ref{di})). Of course, this fact does not spoil our closures at all, but it makes it easier to deal with correlation functions.

\subsection{Coarse-grained velocity}

 The coefficients $A(r)$ and $B(r)$ appearing in (\ref{ab}) are uniquely determined by the system
 \[\left\{\begin{array}{ll}
  2rB'(r)-6B(r)=-rG(r)&\\
  3A(r)+B(r)=G(r)&,
 \end{array}\right.\]
 with $B(0)=0$ and
 \begin{eqnarray*}
  G(r)\!\!&\!\!=\!\!&\!\!-\frac{2^{\xi}144D_0l^{\xi}}{(4+\xi)(6+\xi)}+\frac{9D_0}{2(2+\xi)(4+\xi)l^6}\left\{-\frac{l}{5+\xi}\left[R_+^{5+\xi}-R|R_-|^{4+\xi}\right]\right.\\
  &&-\frac{l^2}{(5+\xi)r}\left[R_+^{5+\xi}+R|R_-|^{5+\xi}-2r^{5+\xi}\right]+\frac{l}{(6+\xi)r}\left[R_+^{6+\xi}-|R_-|^{6+\xi}\right]\\
  &&+\frac{1}{(5+\xi)(6+\xi)}\left[R_+^{6+\xi}+|R_-|^{6+\xi}-2r^{6+\xi}\right]+\frac{l}{(5+\xi)(6+\xi)r}\times\\
  &&\times\left[R_+^{6+\xi}-|R_-|^{6+\xi}\right]\left.-\frac{1}{(5+\xi)(7+\xi)r}\left[R_+^{7+\xi}+R|R_-|^{7+\xi}-2r^{7+\xi}\right]\right\}\;.
 \end{eqnarray*}
 A series expansion in $l/r$ gives
 \[A(r)=D_0r^{\xi}\left[(2+\xi)-\frac{2^{\xi}48}{(4+\xi)(6+\xi)}\left(\frac{l}{r}\right)^{\xi}+\frac{1}{5}\xi^2(3+\xi)\left(\frac{l}{r}\right)^2+{\rm O}\left(\frac{l}{r}\right)^4\right]\]
 and
 \[B(r)=D_0r^{\xi}\left[-\xi+\frac{1}{5}\xi(3+\xi)(2-\xi)\left(\frac{l}{r}\right)^2+{\rm O}\left(\frac{l}{r}\right)^4\right]\;.\]

\section{Limit cases: purely-diffusive or smooth flow} \label{sec:pdsf}

 The values $\xi=0$ and $\xi=2$ have been excluded from our analysis up to now, because they represent two limit cases: a purely diffusive flow and a smooth flow, respectively. It is however interesting to analyse them, as in the former case an exact closure can be found, and in the latter a logarithmic law arises.

\subsection{The case $\xi=0$}

 In the case $\xi=0$ the definition of $\eta$ (\ref{eta}) is meaningless, but physically it corresponds to a diffusive range extending to infinity. This can be understood very simply noticing that the velocity field reduces to a white noise, as its second-order moment (\ref{Dmunu}) takes a diagonal form completely independent from the separation: $D_{\mu\nu}^{(\bm{v})}(\bm{r})=(d-1)D_0\delta_{\mu\nu}$.\\
 Consequently, the overall effect of the advective term is barely of diffusive type and consists in an addition of the quantity $(d-1)D_0$ to the original contribution $2\kappa_0$ in (\ref{c2t:eq}). As there is no more need to split the integration interval in (\ref{c2t:sol}), the correlation is exactly expressed by
 \[C_2^{(\theta)}(r)=c-kr^2\qquad\textrm{for }r<L\;,\]
 with
 \[c=\frac{F_0L^2}{2(d-2)[2\kappa_0+(d-1)D_0]}=\langle\theta^2\rangle\qquad\textrm{for }d\ne2\]
 and
 \[k=\frac{F_0}{2d[2\kappa_0+(d-1)D_0]}\;.\]
 By definition, an exact calculation in 3-D yields:
 \[C_2^{(\tilde{\theta})}(r)=c-kr^2\left[1+\frac{6}{5}\left(\frac{l}{r}\right)^2\right]=c'-kr^2\;,\]
 with
 \[c'=c-\frac{F_0l^2}{10(\kappa_0+D_0)}=\langle\tilde{\theta}^2\rangle\;.\]
 In this case, the structure functions of the coarse-grained field and of the original one turn out to be identical: $S_2^{(\tilde{\theta})}(r)=2kr^2=S_2^{(\theta)}(r)$.\\
 The interesting point is that CED closure (\ref{CED}) is now exact also for the correlation function, as it happened in (\ref{keff}) only for the mean value. This is in accordance with the vanishing of the last term on the right-hand side of (\ref{DED2}). From the analytical point of view, the exactness of the eddy-diffusivity closure is due to the vanishing of the second-order structure function of the coarse-grained velocity, $D_{\mu\nu}^{(\tilde{\bm{v}})}(\bm{r})=0$, which imposes an exact balancing between the forcing and the diffusive terms in (\ref{CED:eq}).

\subsection{The case $\xi=2$}

 If $\xi=2$, the second-order spatial increments of the velocity (\ref{Dmunu}) scale with $r^2$, so $\bm{v}$ is a differentiable field. The passive-scalar correlation is exactly given by
 \begin{equation} \label{csi2}
  C_2^{(\theta)}(r)=c-k\ln\frac{r^2+\eta^2}{L^2+\eta^2}\qquad\textrm{for }r<L\;,
 \end{equation}
 where
 \[k=\frac{F_0}{6(d-1)D_0}\]
 and
 \[c=-\frac{F_0}{d^2(d-1)D_0}{}_2{\rm F}_1\left(1,\frac{d}{2};1+\frac{d}{2};\frac{2\kappa_0}{(d-1)D_0L^2}\right)\;,\]
 ${}_2{\rm F}_1$ being the hypergeometric function.\\
 Exploiting its definition, for $d=3$, the coarse-grained scalar correlation shows a second-order correction in $l/r$ with respect to its fully-resolved corresponding value (\ref{csi2}),
 \begin{equation} \label{xi2}
  C_2^{(\tilde{\theta})}(r)=C_2^{(\theta)}-\frac{F_0}{30D_0}\left(\frac{l}{r}\right)^2+{\rm O}\left(\frac{l}{r}\right)^4\;,
 \end{equation}
 while no difference exists for the velocity structure function: $D_{\mu\nu}^{(\bm{v})}=D_{\mu\nu}^{(\tilde{\bm{v}})}$.\\
 The key point is that ${\rm O}(l/r)^{\xi}$ and ${\rm O}(l/r)^2$ obviously coincide, so CED closure cannot be introduced by itself because it is intimately entangled with the improved closure (\ref{DED2}). The latter captures the second-order correction in (\ref{xi2}) correctly.

\end{document}